\documentclass[aps, prl, twocolumn, superscriptaddress]{revtex4-1}

\usepackage[tbtags,intlimits]{amsmath}

\usepackage[pdftex]{graphicx}
\usepackage{epsfig}
\usepackage{amsmath}
\usepackage{grffile} 
\usepackage{subfigure}

\usepackage{amsmath}
\usepackage{graphics}
\usepackage{epsfig}
\usepackage{graphicx}

\usepackage{ulem}

\usepackage{color}

\usepackage[squaren]{SIunits}

\let\vec\boldsymbol

\newcommand{\mean}[1]{\langle{#1}\rangle}

\newif\ifWC
\WCfalse

\begin{document}

\title{Goldilocks Zone for Enhanced Ionization in Strong Laser Fields}

\author{M. M\"oller}
\affiliation{Institute of Optics and Quantum Electronics, Abbe Center of Photonics, Friedrich Schiller University Jena, Max-Wien-Platz 1, 07743 Jena, Germany}
\affiliation{Helmholtz Institut Jena, Fr\"obelstieg 3, 07743 Jena, Germany}

\author{A. M. Sayler}
\affiliation{Institute of Optics and Quantum Electronics, Abbe Center of Photonics, Friedrich Schiller University Jena, Max-Wien-Platz 1, 07743 Jena, Germany}
\affiliation{Helmholtz Institut Jena, Fr\"obelstieg 3, 07743 Jena, Germany}
\email[corresponding author]{sayler@uni-jena.de}

\author{P. Wustelt}
\affiliation{Institute of Optics and Quantum Electronics, Abbe Center of Photonics, Friedrich Schiller University Jena, Max-Wien-Platz 1, 07743 Jena, Germany}
\affiliation{Helmholtz Institut Jena, Fr\"obelstieg 3, 07743 Jena, Germany}

\author{L. Yue}
\affiliation{Institute of Physical Chemistry and Abbe Center of Photonics, Friedrich Schiller University Jena, Helmholtzweg 4, 07743 Jena, Germany}

\author{S. Gr\"afe}
\affiliation{Institute of Physical Chemistry and Abbe Center of Photonics, Friedrich Schiller University Jena, Helmholtzweg 4, 07743 Jena, Germany}

\author{G. G. Paulus}
\affiliation{Institute of Optics and Quantum Electronics, Abbe Center of Photonics, Friedrich Schiller University Jena, Max-Wien-Platz 1, 07743 Jena, Germany}
\affiliation{Helmholtz Institut Jena, Fr\"obelstieg 3, 07743 Jena, Germany}

\begin{abstract}
Utilizing a benchmark measurement of laser-induced ionization of an H$_2^+$ molecular ion beam target at infrared wavelength around 2 $\mu$m, we show that the characteristic two-peak structure predicted for laser-induced enhanced ionization of H$_2^+$ and diatomic molecules in general, is a phenomenon which is confined to a small laser parameter space --- a Goldilocks Zone. 
Further, we control the effect experimentally and measure its imprint on the electron momentum. We replicate the behavior with simulations, which reproduce the measured kinetic-energy release as well as the correlated-electron spectra. Based on this, a model, which both maps out the Goldilocks Zone and illustrates why enhanced ionization has proven so elusive in H$_2^+$, is derived.
\end{abstract}


\maketitle

Since its first complete quantum description in the 1920's \cite{H2p1927}, the H$_2^+$ bond has served as the prototype for all molecular systems \cite{Ibrahim.JPB.2018}.  This is particularly true for the attosecond dynamics of molecular bonding in strong fields, where the insights gained from H$_2^+$ have served as a foundation for the understanding of more complex bonds. For example, the concepts of bond hardening, 
bond softening,
above-threshold dissociation
and above-threshold ionization 
were first determined from H$_2^+$ and are now ubiquitous in the descriptions of laser-induced molecular dynamics \cite{Ibrahim.JPB.2018}. 

However, another foundational process --- enhanced ionization (EI) --- continues to be evasive and contentious \cite{Ben-Itzhak2008a, Xu2015}, partially due to the difficulty of direct measurements of  H$_2^+$. The EI process was predicted by early time-dependent Schr\"odinger equation (TDSE) calculations for H$_2^+$ with fixed internuclear distances and showed that ionization, i.e.~H$_2^+\rightarrow p^++p^++e^-$,  is enhanced for specific internuclear distances \cite{Zuo1995a}. 
Since, EI has frequently been used as one of the processes invoked to explain and predict laser-induced molecular dynamics for small and more complex molecules alike \cite{Normand.PRA.1996,Hishikawa.PRL.1999,Liekhus2015,Jiang2010,Xie.PRA.2014,Cornaggia.JPB.2016,Liu.PRL.2017}. 

\begin{figure}
\centering
\includegraphics[trim = 0mm 0mm 0mm 0mm, width=0.99 \columnwidth]{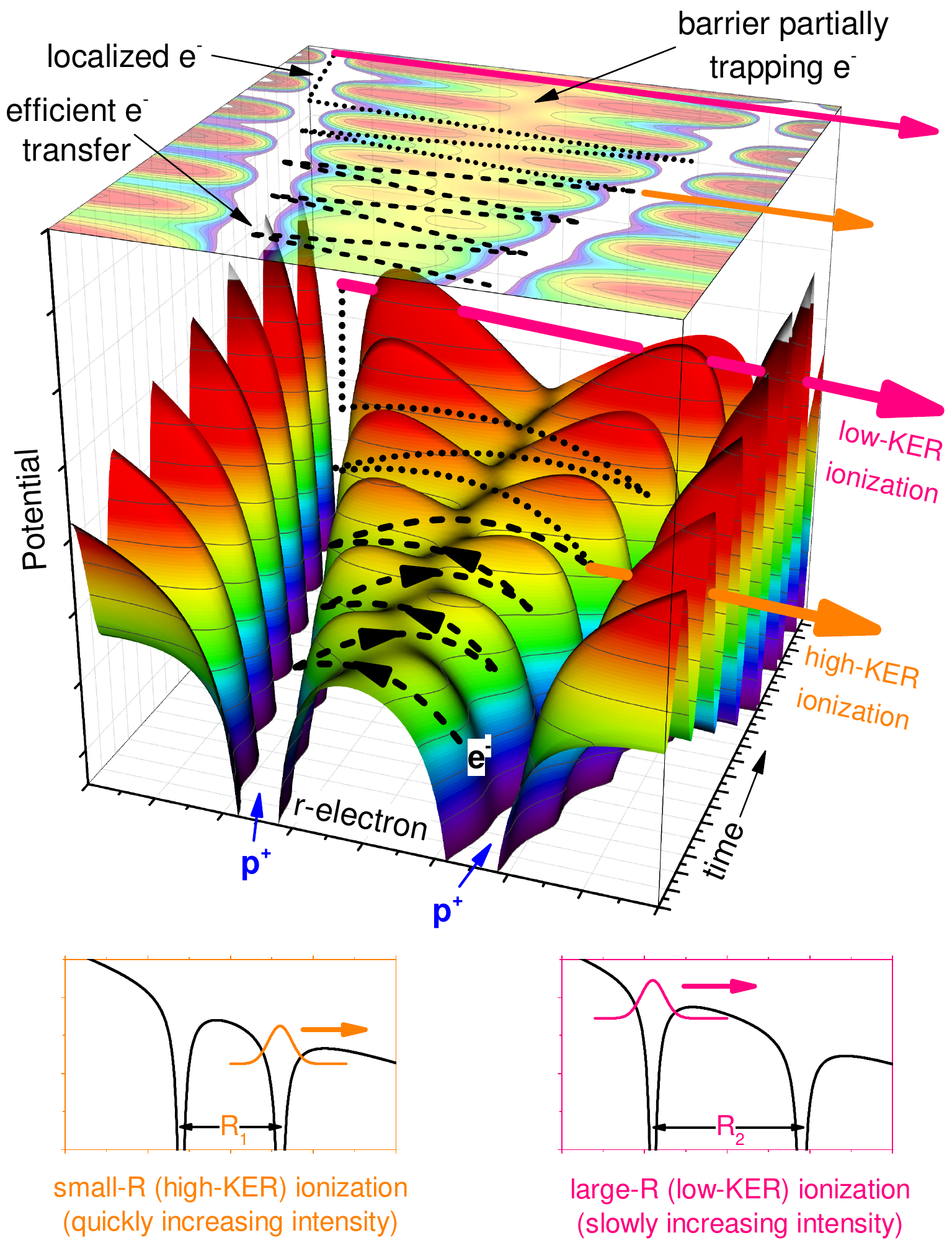}
\caption{Schematic of enhanced ionization (EI). (top) Time- and $r$-dependent potential with projection. Stretching of the internuclear distance, $R$, is initiated early in the pulse and the electron, $e^-$, is driven back and forth between the two nuclei by the laser field (dashed arrows). For electric fields that increase quickly in time, after a few cycles the electron can ionize from the lower potential (bottom left) resulting in large kinetic-energy release (KER). For electric fields that increase slowly in time, the field is not large enough to ionize at small $R$ and the electron continues (dotted arrows) to oscillate, as the molecule stretches, until the internuclear distance is large enough to create an internuclear potential barrier (bottom right) partially localizing the electron and facilitating ionization from the upper well. This enhances the ionization probability resulting in small KER. }
\label{fig:1}
\end{figure}

In a common picture of EI in H$_2^+$, depicted in Fig.~\ref{fig:1}, there are two distinct mechanisms leading to ionization. If the nuclear separation, $R$, is relatively small and the laser field is relatively large, then the electronic wavefunction will follow the laser, moving to the downhill side of the potential with each optical half cycle and tunnel out with sufficient laser intensity around the peaks of the field. Alternatively, if the laser field ramps up slowly, the molecule stretches and the potential barrier between the two protons grows. This can effectively trap a portion of the electron wavepacket on the uphill proton and allow ionization from the upper potential over the deformed inner barrier \cite{Liu.PRL.2017}. Thus, ionization is enhanced near $R$-values corresponding to the aforementioned cases and a characteristic double peak structure should emerge in the $R$- or KER-dependent electron yield \cite{Zuo1995a}.

Although this is a logical and straightforward explanation and the double-peak structure is obvious in the calculated static-field ionization rate of fixed-nuclei H$_2^+$ molecules \cite{Plummer1996,Plummer1997}, 
many 
studies of H$_2^+$ over the past two decades could not clearly identify EI, 
which has fueled the debate over the relevance of the concept \cite{Takemoto2010,Odenweller2011,Silva2013a,Xu2015,Yue2016b,Xu2017}.
Moreover, experimental effects; such as the initial vibrational state distribution of a H$_2^+$ target, the imprint of the prerequisite ionization of a H$_2$ target, depletion, intensity-volume effects, conversion from the measured KER to the inferred $R$-values and the coupled electron-nuclear dynamics; make the process and interpretation much more complex \cite{Ben-Itzhak2008a, Xu2015}. 
Measurements, which use neutral H$_2$ as a target, often leave uncertainty about which observed effects are, at least partially, due to prerequisite ionization.
For example, indications of EI have only been clearly seen when creating a traveling nuclear wavepacket from a ionization from a neutral target, $\textrm{H}_2 \rightarrow \textrm{H}_2^+ + e^-$, and probing the resulting dissociative state with time-delayed few-cycle pulses \cite{Xu2015}. However, this leaves many question about the prevalence and importance of the phenomenon for typical laser fields.

To settle these long-standing questions and clearly demonstrate the EI effect in H$_2^+$, here, we implement, the first to our knowledge, intensity-dependent measurement of the short-wave infrared (SWIR) laser-induced ionization of a H$_2^+$ molecular-ion-beam target, which captures both the momenta of the nuclear fragments and the correlated electron momentum. In this largely unexplored territory of wavelength and intensity, we conclusively observe the two-peak structure characteristic of EI and are able to control it with the intensity envelope of the laser pulse.  Next, using specifically developed nuclear-wavepacket propagation calculations that include ionization, which are further validated by the  correlated KER and electron momenta spectra, we determine the Goldilocks Zone, i.e.~the limited laser parameter space, where EI is visible. Finally, this data is used to formulate a clear-cut model with great explanatory and predictive power.  

The experimental challenges arise from the dilute ion-beam target and conditions needed to measure the momenta of both protons and the electron  \cite{Ben-Itzhak2005, Rathje2013, Wustelt2015}, see supplemental material for details \footnote{See Supplemental Material for a PDF document further detailing the experimental methods and numerical simulations.}. The electron momentum, $\vec{p_{e}}$, is determined using the sum momentum of the proton, i.e.~without directly detecting the electron.  This requires an extremely high experimental precision and well-collimated ion beam. Further, despite the significantly more complex laser setup needed, we chose to use SWIR pulses to increase the electron momentum, $|\vec{p}_{e}|\propto \lambda$. Although our measurement of the electron momentum is blurred with the momentum distribution that arises from the temperature of the H$_2^+$ molecular ion beam, compared to a direct measurement of the electron momentum \cite{Odenweller2011,Odenweller2014}, our approach reduces the experimental cost and complexity significantly and is, to the best of our knowledge, the first application of this technique to a molecular ion beam laser interaction.

\begin{figure}
\centering
\hspace*{-0.75cm} 
\includegraphics[trim = 0mm 125mm 264mm 0mm, clip, width=0.72 \columnwidth]{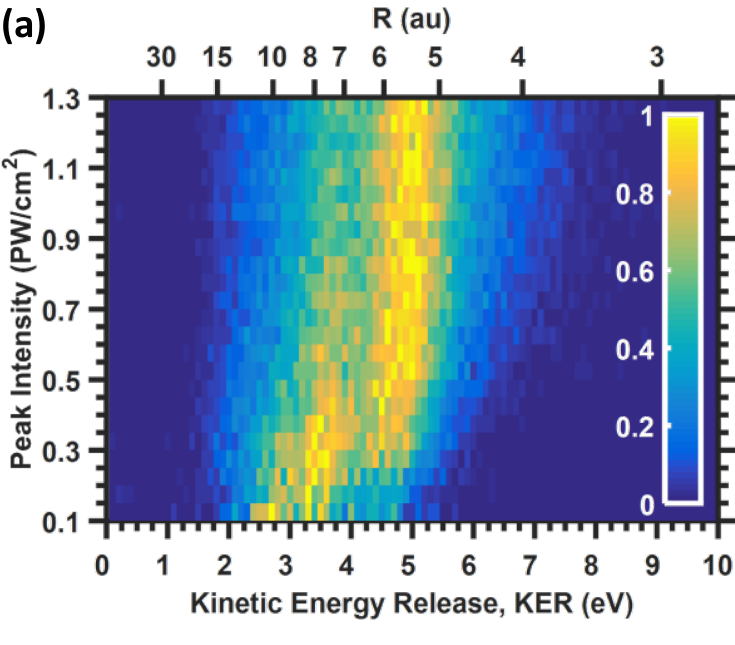}
\includegraphics[trim = 76mm 125mm 180mm 0mm, clip, width=0.8 \columnwidth]{fig2.pdf}
\caption{(a) Measured intensity- and KER-dependent ionization yield of H$_2^+$ for a 65 fs 2 $\mu$m pulse. The shape of the KER spectra is emphasized by normalization of the spectrum within each intensity bin of the 2D plot. (b) Logarithm of the measured joint electron-nuclear energy distribution (JED) for $I_0 \approx (0.75\pm 0.25) \times 10^{15}$ W/cm$^2$. The KER spectrum (grey line) and the mean electron energy are overlaid (orange dots). Note the KER-dependent modulation of the width of the photoelectron spectrum, which has minima where the KER-dependent yield peaks.  See text for details. \ifWC (98) \fi}
\label{fig:2}
\end{figure}

The measured intensity- and KER-dependent laser-induced ionization yield for H$_2^+$ with 65 fs, 2 $\mu$m pulses is shown in Fig.~\ref{fig:2}(a). Here one sees that at low intensity ($I\lesssim$ 0.5 PW/cm$^2$) the yield is peaked near 3.5 eV ($R_2\simeq7.5$ a.u.). For these intensities, the laser field ramps up slowly and the molecule stretches, trapping a portion of the electron wavepacket in the upper potential well, which is then ionized. At high intensity ($I\gtrsim$ 0.3 PW/cm$^2$) the yield is peaked near 5 eV ($R_1\simeq5.5$ a.u.) and slowly increases with intensity. Here, the steep increase in intensity facilitates ionization at smaller $R$s, where the electron wavepacket is effectively transferred to the lower potential well each half cycle. This depletes the dissociative nuclear wavepacket before it reaches $R_2$ thereby reducing the peak at lower KER.
The characteristic double-peak structure we are in search of only occurs in the narrow overlapping transition intensity range ($I\approx$ 0.4$\pm$0.2 PW/cm$^2$), where both processes can occur.

Although this interpretation, based on the measurement of the nuclear fragments, tells us a great deal about the underlying dynamics, it is only half the picture. To gain full access to the dynamics at play, we simultaneously measure the nuclear fragment momenta to produce the joint electron-nuclear energy distribution (JED) shown in Fig.~\ref{fig:2}(b). Here we see that reductions in the width of the electron spectrum are correlated with increases in the yield.
Although calculated JEDs, where the electron and the nuclear dynamics are treated in equal footing \cite{Silva2013a,Yue2013,Yue2016b}, show hints of this behavior, that work typically focuses on diagonal energy-conserving lines, which have also been measured, e.g. by Wu et al.~\cite{Wu2013}. In contrast, here we are focused on the large-scale behavior.

Unlike some of the more complex models of H$_2^+$ ionization, which look at details of the $R$-dependent timing of the liberated electron wave-packet \cite{Takemoto2010,Takemoto2011}, the measurement behavior here has a relatively straightforward qualitative explanation. Namely, in addition to increasing the yield, enhancing the ionization rate at certain $R$s lowers the average intensity required for ionization. This, in turn, reduces the photoelectron energy, which scales with the intensity \cite{Becker2002}. Thus, peaks in the KER-dependent ionization yield should overlap with minima in the width of the correlated electron spectrum as observed. \ifWC (91) \fi

\begin{figure}
\centering
\hspace*{-1.25cm} 
\includegraphics[trim = 0mm 130mm 232mm 20mm, clip, width=0.744 \columnwidth]{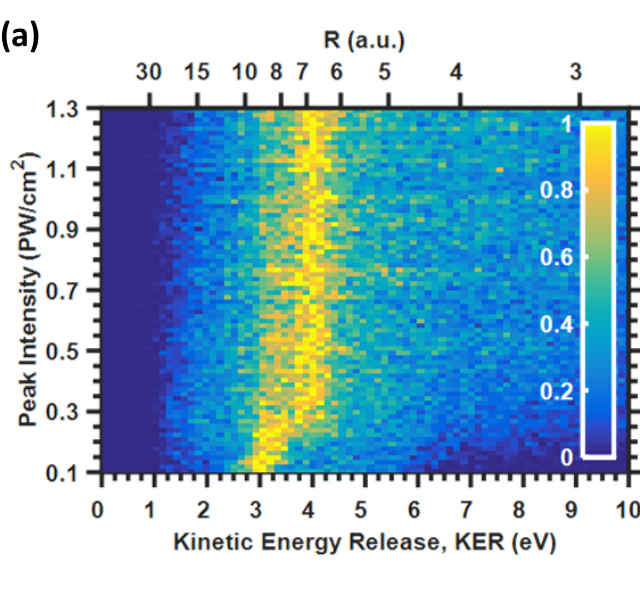}
\includegraphics[trim = 67mm 130mm 160mm 20mm, clip, width=0.8 \columnwidth]{fig3.pdf}
\caption{(a) Calculated intensity- and KER-dependent ionization yield of H$_2^+$ for a 65 fs 2 $\mu$m pulse. The shape of the KER spectra is emphasized by normalization of the spectrum within each intensity bin of the 2D plot and jitter has been added by simulating the measurement statistics. (b) Calculated $\mean{E_{e}}$ and the KER spectrum at a peak intensity of $I_0 = 0.75 \times 10^{15}$ W/cm$^2$. All calculation results are averaged over the intensity distribution in the focal volume as well as over the relevant vibrational states.}
\label{fig:3}
\end{figure}

To further understand the experimental results and extend the control of EI to other  laser parameters, e.g.~wavelength and pulse duration, we implement a two-surface time-dependent Schr\"odinger calculations \cite{Schwendner1997} and augmented them for ionization \cite{Staudte2007} including the correlated electron energy, see supplemental material. Using the measured laser parameters, this results in the spectra shown in Fig.~\ref{fig:3}. Here we see that the enhanced two-surface model is a good qualitative match to the measured data and accurately predicts the double-peak structure for roughly the same narrow intensity range. Moreover, the corresponding KER-dependent electron momentum also follows the measured trend, which confirms that the model is capturing the relevant underlying dynamics. The minor differences between measurement and theory, i.e.~slightly different peak positions, are likely due to several imperfections of the model detailed in the supplemental material.

\begin{figure*}
\centering
\includegraphics[trim = 0mm 30mm 0mm 0mm, clip, width=1.0 \textwidth]{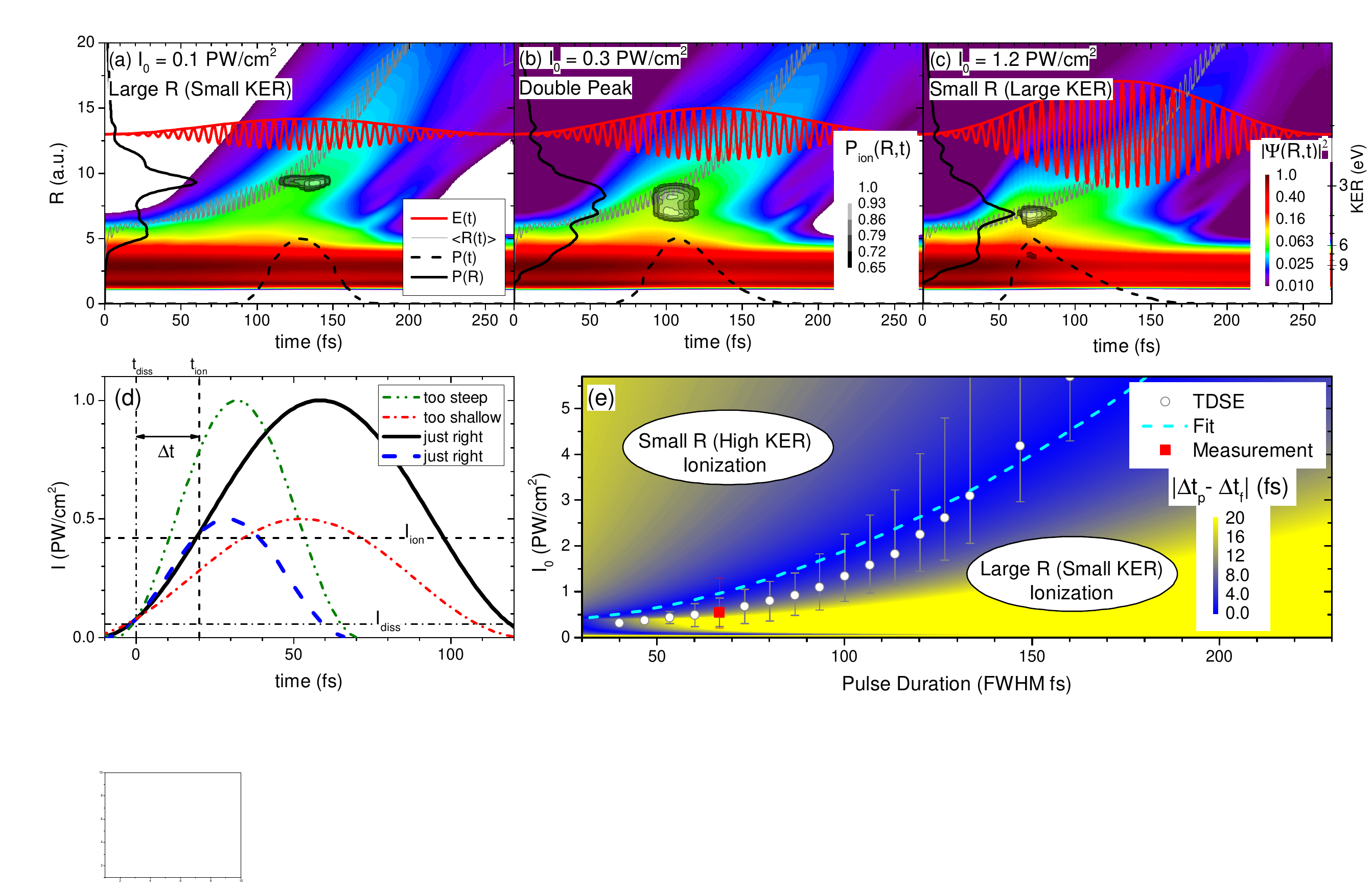}
\caption{(a)--(c) In color, calculated probability density, $|\Psi(R,t)|^2$, as a function of time, $t$, and internuclear distance, $R$, for three different intensities, $I_0 = 0.1$,  $0.29$ and $1.2$ $\textrm{PW/cm}^2$, respectively. This is intensity averaged over the experimental 2D target geometry and comprised of the incoherent sum of vibrational states. The expectation value of $R$, $\langle R \rangle$, is shown as a gray line to guide the eye. The $t$- and $R$-dependent ionization probability are displayed as contours with projections on the bottom and left axes, respectively. For reference, the laser field, including its envelope, is shown in red at the top. Note the legends apply to all three pannels. (d) Examples of pulses, $I(I_0,\textrm{FWHM};t)$, that fulfill the requirements to result in a double peak in the KER-dependent ionization yield (bold lines) and those that do not (thin lines). (e) The position of the measured KER-dependent double peak with the line marking the region where the double peak is visible (red). Additionally,  the positions of the most predominant KER-dependent double peak from the calculations (white circles), with lines marking the region where the double peak is visible. In color, a map of the time delay, $\Delta t_{\textrm{pulse}}(I_{\textrm{diss}},I_{\textrm{ion}};\textrm{FWHM},I_0)$, between $I_{\textrm{diss}}$ and $I_{\textrm{ion}}$ for a given pulse, with respect to the travel time determined from the fit parameters, i.e.~$|\Delta t_{\textrm{pulse}}(I_{\textrm{diss}},I_{\textrm{ion}};\textrm{FWHM},I_0)-\Delta t_{\textrm{fit}}|$.  See text for details.}
\label{fig:4}
\end{figure*}

To identify the dynamics at play we examine 
the calculated nuclear dynamics for the three characteristic situations noted above in Fig.~\ref{fig:4}(a)--(c). In Fig.~\ref{fig:4}(a), when the intensity is \textit{too small}, the leading edge of the laser pulse begins the dissociation process, i.e.~stretching in $R$ of the nuclear wavepacket, and the intensity only becomes sufficient to ionize after the molecule has stretched to $R_2\simeq9$ a.u. (KER $\simeq$ 3 eV).  In Fig.~\ref{fig:4}(c), when the intensity is \textit{too big}, the intensity ramps up quickly enough that ionization depletes the nuclear wavepacket  at $R_1\simeq 6$ a.u. (KER $\simeq$ 4.5 eV) before much stretching can occur. In Fig.~\ref{fig:4}(b), when the intensity is \textit{just right}, the intensity is high enough to ionize near $R_1$, but low enough to allow part of the wavepacket to survive and stretch to $R_2$ before ionization, which results in the double peak shown. 

This leads to an intuitive model for predicting when the characteristic double-peak structure of EI will be visible. Assuming that stretching of the molecule is initiated by the laser field at relatively low intensities, $I_{\textrm{diss}}$, and small internuclear distances, $R_{\textrm{diss}}$, then ionization will occur at a later time, $t_{\textrm{ion}} = t_{\textrm{diss}} + \Delta t$, after the molecule has stretched to $R_{\textrm{ion}}$ and the intensity has increased sufficiently to ionize the molecule, $I_{\textrm{ion}}$.  Further, if there are two preferred internuclear distances for ionization, $R_1^{\textrm{ion}}$ and $R_2^{\textrm{ion}}$, then the double-peaked structure will be lost, if the laser intensity does not ramp up in the very particular way described above. This places constraints on the intensity envelope of the laser pulse. For example, if one wishes to maintain the same timing while increasing the intensity, the pulse length must be increased, see the thick dashed and solid lines in Fig.~\ref{fig:4}(d).

To determine these parameters, we first map out this region with our measurements and wavepacket-propagation calculations. Specifically, to parameterize the double peak structure, we fit a double Gaussian
to the calculated data over a large range of laser parameters. Then we find the ratio of the peaks, $\alpha=A_< / A_>$,
where $A_<$ is the amplitude of the lower of the two peaks and $A_>$ is the the amplitude of the higher of the two peaks. This allows one to determine the Goldilocks Zone where the double peak occurs and to what extent it is visible. The intensity where the maximum value of this ratio, $\alpha_{\textrm{max}}$, occurs is plotted in Fig.~\ref{fig:4}(e) (circles) as a function of the FWHM pulse duration for the calculations at $\lambda=2$ $\mu$m and $\alpha>0.9$ is marked by the bars.

This data can then be used to test the aforementioned model, see supplemental material, and extract the fit parameters from the intensity averaged calculations: $I_{\textrm{diss}}\simeq 5.8\cdot 10^{13}~\textrm{W/cm}^2$, $I_{\textrm{ion}}\simeq 4.2\cdot 10^{14}~\textrm{W/cm}^2$ and $\Delta t \simeq 20~\textrm{fs}$, see line in Fig.~\ref{fig:4}(e). Here the positions of the calculated maximum ratio between the two peaks fit nicely to the simple model and the fitted values are consistent with existing measurements and calculations. Therefore, this remarkably simple model can serve as a guide to control EI by balancing the pulse length and intensity of the laser.  Additionally, the time for the pulse to ramp up from $I_{\textrm{diss}}$ to $I_{\textrm{ion}}$ relative to $\Delta t$ is plotted in false color to illustrate why EI is so elusive, particularly for the short pulses typically used in strong-field physics.

In conclusion, we have measured the intensity- and KER-dependent ionization yield, along with the electron momentum, for the benchmark molecular ion, H$_2^+$, starting directly from a molecular ion beam in the relatively unexplored short-wave infrared (SWIR) regime. We demonstrate that the characteristic double-peak feature of enhanced ionization (EI) can only be observed in a very limited laser parameter space --- the Goldilocks Zone --- where pulse duration and laser intensity are carefully balanced and the interplay between nuclear stretching dynamics and ionization allows for ionization from a broad nuclear wave packet. This directly address a long-standing debate, explains the elusive nature of enhanced ionization, and serves as a guide for how to manipulate laser parameters to coherently control the phenomenon. Moreover, as the behavior of H$_2^+$ serves as the prototype for all molecular systems and EI is generally believed to play a decisive role in more complex systems, this has broad ramifications for strong-field physics in general.

\begin{acknowledgments}
We acknowledge helpful discussions with F.~Grossmann. This work was supported by grants PA730/5 and GR4482/2 of the German Research Foundation (DFG) as well as by laserlab europe.
\end{acknowledgments}


%


\begin{thebibliography}{30}%
\makeatletter
\providecommand \@ifxundefined [1]{%
 \@ifx{#1\undefined}
}%
\providecommand \@ifnum [1]{%
 \ifnum #1\expandafter \@firstoftwo
 \else \expandafter \@secondoftwo
 \fi
}%
\providecommand \@ifx [1]{%
 \ifx #1\expandafter \@firstoftwo
 \else \expandafter \@secondoftwo
 \fi
}%
\providecommand \natexlab [1]{#1}%
\providecommand \enquote  [1]{``#1''}%
\providecommand \bibnamefont  [1]{#1}%
\providecommand \bibfnamefont [1]{#1}%
\providecommand \citenamefont [1]{#1}%
\providecommand \href@noop [0]{\@secondoftwo}%
\providecommand \href [0]{\begingroup \@sanitize@url \@href}%
\providecommand \@href[1]{\@@startlink{#1}\@@href}%
\providecommand \@@href[1]{\endgroup#1\@@endlink}%
\providecommand \@sanitize@url [0]{\catcode `\\12\catcode `\$12\catcode
  `\&12\catcode `\#12\catcode `\^12\catcode `\_12\catcode `\%12\relax}%
\providecommand \@@startlink[1]{}%
\providecommand \@@endlink[0]{}%
\providecommand \url  [0]{\begingroup\@sanitize@url \@url }%
\providecommand \@url [1]{\endgroup\@href {#1}{\urlprefix }}%
\providecommand \urlprefix  [0]{URL }%
\providecommand \Eprint [0]{\href }%
\providecommand \doibase [0]{http://dx.doi.org/}%
\providecommand \selectlanguage [0]{\@gobble}%
\providecommand \bibinfo  [0]{\@secondoftwo}%
\providecommand \bibfield  [0]{\@secondoftwo}%
\providecommand \translation [1]{[#1]}%
\providecommand \BibitemOpen [0]{}%
\providecommand \bibitemStop [0]{}%
\providecommand \bibitemNoStop [0]{.\EOS\space}%
\providecommand \EOS [0]{\spacefactor3000\relax}%
\providecommand \BibitemShut  [1]{\csname bibitem#1\endcsname}%
\let\auto@bib@innerbib\@empty
\bibitem [{\citenamefont {Burrau}(1927)}]{H2p1927}%
  \BibitemOpen
  \bibfield  {author} {\bibinfo {author} {\bibfnamefont {{\O{}}.}~\bibnamefont
  {Burrau}},\ }\href {https://doi.org/10.1007/BF01504875} {\bibfield  {journal}
  {\bibinfo  {journal} {The Science of Nature (Naturwissenschaften)}\ }\textbf
  {\bibinfo {volume} {15}},\ \bibinfo {pages} {16} (\bibinfo {year}
  {1927})}\BibitemShut {NoStop}%
\bibitem [{\citenamefont {Ibrahim}\ \emph {et~al.}(2018)\citenamefont
  {Ibrahim}, \citenamefont {Lefebvre}, \citenamefont {Bandrauk}, \citenamefont
  {Staudte},\ and\ \citenamefont {L\'egar\'e}}]{Ibrahim.JPB.2018}%
  \BibitemOpen
  \bibfield  {author} {\bibinfo {author} {\bibfnamefont {H.}~\bibnamefont
  {Ibrahim}}, \bibinfo {author} {\bibfnamefont {C.}~\bibnamefont {Lefebvre}},
  \bibinfo {author} {\bibfnamefont {A.~D.}\ \bibnamefont {Bandrauk}}, \bibinfo
  {author} {\bibfnamefont {A.}~\bibnamefont {Staudte}}, \ and\ \bibinfo
  {author} {\bibfnamefont {F.}~\bibnamefont {L\'egar\'e}},\ }\href
  {http://stacks.iop.org/0953-4075/51/i=4/a=042002} {\bibfield  {journal}
  {\bibinfo  {journal} {J. Phys. B: At. Mol. Opt. Phys.}\ }\textbf {\bibinfo
  {volume} {51}},\ \bibinfo {pages} {042002} (\bibinfo {year}
  {2018})}\BibitemShut {NoStop}%
\bibitem [{\citenamefont {Ben-Itzhak}\ \emph {et~al.}(2008)\citenamefont
  {Ben-Itzhak}, \citenamefont {Wang}, \citenamefont {Sayler}, \citenamefont
  {Carnes}, \citenamefont {Leonard}, \citenamefont {Esry}, \citenamefont
  {Alnaser}, \citenamefont {Ulrich}, \citenamefont {Tong}, \citenamefont
  {Litvinyuk}, \citenamefont {Maharjan}, \citenamefont {Ranitovic},
  \citenamefont {Osipov}, \citenamefont {Ghimire}, \citenamefont {Chang},\ and\
  \citenamefont {Cocke}}]{Ben-Itzhak2008a}%
  \BibitemOpen
  \bibfield  {author} {\bibinfo {author} {\bibfnamefont {I.}~\bibnamefont
  {Ben-Itzhak}}, \bibinfo {author} {\bibfnamefont {P.~Q.}\ \bibnamefont
  {Wang}}, \bibinfo {author} {\bibfnamefont {A.~M.}\ \bibnamefont {Sayler}},
  \bibinfo {author} {\bibfnamefont {K.~D.}\ \bibnamefont {Carnes}}, \bibinfo
  {author} {\bibfnamefont {M.}~\bibnamefont {Leonard}}, \bibinfo {author}
  {\bibfnamefont {B.~D.}\ \bibnamefont {Esry}}, \bibinfo {author}
  {\bibfnamefont {A.~S.}\ \bibnamefont {Alnaser}}, \bibinfo {author}
  {\bibfnamefont {B.}~\bibnamefont {Ulrich}}, \bibinfo {author} {\bibfnamefont
  {X.~M.}\ \bibnamefont {Tong}}, \bibinfo {author} {\bibfnamefont {I.~V.}\
  \bibnamefont {Litvinyuk}}, \bibinfo {author} {\bibfnamefont {C.~M.}\
  \bibnamefont {Maharjan}}, \bibinfo {author} {\bibfnamefont {P.}~\bibnamefont
  {Ranitovic}}, \bibinfo {author} {\bibfnamefont {T.}~\bibnamefont {Osipov}},
  \bibinfo {author} {\bibfnamefont {S.}~\bibnamefont {Ghimire}}, \bibinfo
  {author} {\bibfnamefont {Z.}~\bibnamefont {Chang}}, \ and\ \bibinfo {author}
  {\bibfnamefont {C.~L.}\ \bibnamefont {Cocke}},\ }\href {\doibase
  10.1103/PhysRevA.78.063419} {\bibfield  {journal} {\bibinfo  {journal} {Phys.
  Rev. A}\ }\textbf {\bibinfo {volume} {78}},\ \bibinfo {pages} {063419}
  (\bibinfo {year} {2008})}\BibitemShut {NoStop}%
\bibitem [{\citenamefont {Xu}\ \emph {et~al.}(2015)\citenamefont {Xu},
  \citenamefont {He}, \citenamefont {Kielpinski}, \citenamefont {Sang},\ and\
  \citenamefont {Litvinyuk}}]{Xu2015}%
  \BibitemOpen
  \bibfield  {author} {\bibinfo {author} {\bibfnamefont {H.}~\bibnamefont
  {Xu}}, \bibinfo {author} {\bibfnamefont {F.}~\bibnamefont {He}}, \bibinfo
  {author} {\bibfnamefont {D.}~\bibnamefont {Kielpinski}}, \bibinfo {author}
  {\bibfnamefont {R.~T.}\ \bibnamefont {Sang}}, \ and\ \bibinfo {author}
  {\bibfnamefont {I.~V.}\ \bibnamefont {Litvinyuk}},\ }\href
  {http://dx.doi.org/10.1038/srep13527} {\bibfield  {journal} {\bibinfo
  {journal} {Sci. Rep.}\ }\textbf {\bibinfo {volume} {5}},\ \bibinfo {pages}
  {1} (\bibinfo {year} {2015})}\BibitemShut {NoStop}%
\bibitem [{\citenamefont {Zuo}\ and\ \citenamefont
  {Bandrauk}(1995)}]{Zuo1995a}%
  \BibitemOpen
  \bibfield  {author} {\bibinfo {author} {\bibfnamefont {T.}~\bibnamefont
  {Zuo}}\ and\ \bibinfo {author} {\bibfnamefont {A.~D.}\ \bibnamefont
  {Bandrauk}},\ }\href@noop {} {\bibfield  {journal} {\bibinfo  {journal}
  {Phys. Rev. A}\ }\textbf {\bibinfo {volume} {52}},\ \bibinfo {pages} {2511}
  (\bibinfo {year} {1995})}\BibitemShut {NoStop}%
\bibitem [{\citenamefont {Normand}\ and\ \citenamefont
  {Schmidt}(1996)}]{Normand.PRA.1996}%
  \BibitemOpen
  \bibfield  {author} {\bibinfo {author} {\bibfnamefont {D.}~\bibnamefont
  {Normand}}\ and\ \bibinfo {author} {\bibfnamefont {M.}~\bibnamefont
  {Schmidt}},\ }\href {\doibase 10.1103/PhysRevA.53.R1958} {\bibfield
  {journal} {\bibinfo  {journal} {Phys. Rev. A}\ }\textbf {\bibinfo {volume}
  {53}},\ \bibinfo {pages} {R1958} (\bibinfo {year} {1996})}\BibitemShut
  {NoStop}%
\bibitem [{\citenamefont {Hishikawa}\ \emph {et~al.}(1999)\citenamefont
  {Hishikawa}, \citenamefont {Iwamae},\ and\ \citenamefont
  {Yamanouchi}}]{Hishikawa.PRL.1999}%
  \BibitemOpen
  \bibfield  {author} {\bibinfo {author} {\bibfnamefont {A.}~\bibnamefont
  {Hishikawa}}, \bibinfo {author} {\bibfnamefont {A.}~\bibnamefont {Iwamae}}, \
  and\ \bibinfo {author} {\bibfnamefont {K.}~\bibnamefont {Yamanouchi}},\
  }\href {\doibase 10.1103/PhysRevLett.83.1127} {\bibfield  {journal} {\bibinfo
   {journal} {Phys. Rev. Lett.}\ }\textbf {\bibinfo {volume} {83}},\ \bibinfo
  {pages} {1127} (\bibinfo {year} {1999})}\BibitemShut {NoStop}%
\bibitem [{\citenamefont {Liekhus-Schmaltz}\ \emph {et~al.}(2015)\citenamefont
  {Liekhus-Schmaltz}, \citenamefont {Tenney}, \citenamefont {Osipov},
  \citenamefont {Sanchez-Gonzalez}, \citenamefont {Berrah}, \citenamefont
  {Boll}, \citenamefont {Bomme}, \citenamefont {Bostedt}, \citenamefont
  {Bozek}, \citenamefont {Carron}, \citenamefont {Coffee}, \citenamefont
  {Devin}, \citenamefont {Erk}, \citenamefont {Ferguson}, \citenamefont
  {Field}, \citenamefont {Foucar}, \citenamefont {Frasinski}, \citenamefont
  {Glownia}, \citenamefont {Gühr}, \citenamefont {Kamalov}, \citenamefont
  {Krzywinski}, \citenamefont {Li}, \citenamefont {Marangos}, \citenamefont
  {Martinez}, \citenamefont {McFarland}, \citenamefont {Miyabe}, \citenamefont
  {Murphy}, \citenamefont {Natan}, \citenamefont {Rolles}, \citenamefont
  {Rudenko}, \citenamefont {Siano}, \citenamefont {Simpson}, \citenamefont
  {Spector}, \citenamefont {Swiggers}, \citenamefont {Walke}, \citenamefont
  {Wang}, \citenamefont {Weber}, \citenamefont {Bucksbaum},\ and\ \citenamefont
  {Petrovic}}]{Liekhus2015}%
  \BibitemOpen
  \bibfield  {author} {\bibinfo {author} {\bibfnamefont {C.~E.}\ \bibnamefont
  {Liekhus-Schmaltz}}, \bibinfo {author} {\bibfnamefont {I.}~\bibnamefont
  {Tenney}}, \bibinfo {author} {\bibfnamefont {T.}~\bibnamefont {Osipov}},
  \bibinfo {author} {\bibfnamefont {A.}~\bibnamefont {Sanchez-Gonzalez}},
  \bibinfo {author} {\bibfnamefont {N.}~\bibnamefont {Berrah}}, \bibinfo
  {author} {\bibfnamefont {R.}~\bibnamefont {Boll}}, \bibinfo {author}
  {\bibfnamefont {C.}~\bibnamefont {Bomme}}, \bibinfo {author} {\bibfnamefont
  {C.}~\bibnamefont {Bostedt}}, \bibinfo {author} {\bibfnamefont {J.~D.}\
  \bibnamefont {Bozek}}, \bibinfo {author} {\bibfnamefont {S.}~\bibnamefont
  {Carron}}, \bibinfo {author} {\bibfnamefont {R.}~\bibnamefont {Coffee}},
  \bibinfo {author} {\bibfnamefont {J.}~\bibnamefont {Devin}}, \bibinfo
  {author} {\bibfnamefont {B.}~\bibnamefont {Erk}}, \bibinfo {author}
  {\bibfnamefont {K.~R.}\ \bibnamefont {Ferguson}}, \bibinfo {author}
  {\bibfnamefont {R.~W.}\ \bibnamefont {Field}}, \bibinfo {author}
  {\bibfnamefont {L.}~\bibnamefont {Foucar}}, \bibinfo {author} {\bibfnamefont
  {L.~J.}\ \bibnamefont {Frasinski}}, \bibinfo {author} {\bibfnamefont {J.~M.}\
  \bibnamefont {Glownia}}, \bibinfo {author} {\bibfnamefont {M.}~\bibnamefont
  {Gühr}}, \bibinfo {author} {\bibfnamefont {A.}~\bibnamefont {Kamalov}},
  \bibinfo {author} {\bibfnamefont {J.}~\bibnamefont {Krzywinski}}, \bibinfo
  {author} {\bibfnamefont {H.}~\bibnamefont {Li}}, \bibinfo {author}
  {\bibfnamefont {J.~P.}\ \bibnamefont {Marangos}}, \bibinfo {author}
  {\bibfnamefont {T.~J.}\ \bibnamefont {Martinez}}, \bibinfo {author}
  {\bibfnamefont {B.~K.}\ \bibnamefont {McFarland}}, \bibinfo {author}
  {\bibfnamefont {S.}~\bibnamefont {Miyabe}}, \bibinfo {author} {\bibfnamefont
  {B.}~\bibnamefont {Murphy}}, \bibinfo {author} {\bibfnamefont
  {A.}~\bibnamefont {Natan}}, \bibinfo {author} {\bibfnamefont
  {D.}~\bibnamefont {Rolles}}, \bibinfo {author} {\bibfnamefont
  {A.}~\bibnamefont {Rudenko}}, \bibinfo {author} {\bibfnamefont
  {M.}~\bibnamefont {Siano}}, \bibinfo {author} {\bibfnamefont {E.~R.}\
  \bibnamefont {Simpson}}, \bibinfo {author} {\bibfnamefont {L.}~\bibnamefont
  {Spector}}, \bibinfo {author} {\bibfnamefont {M.}~\bibnamefont {Swiggers}},
  \bibinfo {author} {\bibfnamefont {D.}~\bibnamefont {Walke}}, \bibinfo
  {author} {\bibfnamefont {S.}~\bibnamefont {Wang}}, \bibinfo {author}
  {\bibfnamefont {T.}~\bibnamefont {Weber}}, \bibinfo {author} {\bibfnamefont
  {P.~H.}\ \bibnamefont {Bucksbaum}}, \ and\ \bibinfo {author} {\bibfnamefont
  {V.~S.}\ \bibnamefont {Petrovic}},\ }\href {\doibase 10.1038/ncomms9199}
  {\bibfield  {journal} {\bibinfo  {journal} {Nature Comm.}\ }\textbf {\bibinfo
  {volume} {6}},\ \bibinfo {pages} {8199} (\bibinfo {year} {2015})}\BibitemShut
  {NoStop}%
\bibitem [{\citenamefont {Jiang}\ \emph {et~al.}(2010)\citenamefont {Jiang},
  \citenamefont {Rudenko}, \citenamefont {Herrwerth}, \citenamefont {Foucar},
  \citenamefont {Kurka}, \citenamefont {K\"uhnel}, \citenamefont {Lezius},
  \citenamefont {Kling}, \citenamefont {van Tilborg}, \citenamefont {Belkacem},
  \citenamefont {Ueda}, \citenamefont {D\"usterer}, \citenamefont {Treusch},
  \citenamefont {Schr\"oter}, \citenamefont {Moshammer},\ and\ \citenamefont
  {Ullrich}}]{Jiang2010}%
  \BibitemOpen
  \bibfield  {author} {\bibinfo {author} {\bibfnamefont {Y.~H.}\ \bibnamefont
  {Jiang}}, \bibinfo {author} {\bibfnamefont {A.}~\bibnamefont {Rudenko}},
  \bibinfo {author} {\bibfnamefont {O.}~\bibnamefont {Herrwerth}}, \bibinfo
  {author} {\bibfnamefont {L.}~\bibnamefont {Foucar}}, \bibinfo {author}
  {\bibfnamefont {M.}~\bibnamefont {Kurka}}, \bibinfo {author} {\bibfnamefont
  {K.~U.}\ \bibnamefont {K\"uhnel}}, \bibinfo {author} {\bibfnamefont
  {M.}~\bibnamefont {Lezius}}, \bibinfo {author} {\bibfnamefont {M.~F.}\
  \bibnamefont {Kling}}, \bibinfo {author} {\bibfnamefont {J.}~\bibnamefont
  {van Tilborg}}, \bibinfo {author} {\bibfnamefont {A.}~\bibnamefont
  {Belkacem}}, \bibinfo {author} {\bibfnamefont {K.}~\bibnamefont {Ueda}},
  \bibinfo {author} {\bibfnamefont {S.}~\bibnamefont {D\"usterer}}, \bibinfo
  {author} {\bibfnamefont {R.}~\bibnamefont {Treusch}}, \bibinfo {author}
  {\bibfnamefont {C.~D.}\ \bibnamefont {Schr\"oter}}, \bibinfo {author}
  {\bibfnamefont {R.}~\bibnamefont {Moshammer}}, \ and\ \bibinfo {author}
  {\bibfnamefont {J.}~\bibnamefont {Ullrich}},\ }\href {\doibase
  10.1103/PhysRevLett.105.263002} {\bibfield  {journal} {\bibinfo  {journal}
  {Phys. Rev. Lett.}\ }\textbf {\bibinfo {volume} {105}},\ \bibinfo {pages}
  {263002} (\bibinfo {year} {2010})}\BibitemShut {NoStop}%
\bibitem [{\citenamefont {Xie}\ \emph {et~al.}(2014)\citenamefont {Xie},
  \citenamefont {Roither}, \citenamefont {Sch\"offler}, \citenamefont {Xu},
  \citenamefont {Bubin}, \citenamefont {L\"otstedt}, \citenamefont
  {Erattuphuza}, \citenamefont {Iwasaki}, \citenamefont {Kartashov},
  \citenamefont {Varga}, \citenamefont {G.~Paulus}, \citenamefont
  {Baltu\ifmmode~\check{s}\else \v{s}\fi{}ka}, \citenamefont {Yamanouchi},\
  and\ \citenamefont {Kitzler}}]{Xie.PRA.2014}%
  \BibitemOpen
  \bibfield  {author} {\bibinfo {author} {\bibfnamefont {X.}~\bibnamefont
  {Xie}}, \bibinfo {author} {\bibfnamefont {S.}~\bibnamefont {Roither}},
  \bibinfo {author} {\bibfnamefont {M.}~\bibnamefont {Sch\"offler}}, \bibinfo
  {author} {\bibfnamefont {H.}~\bibnamefont {Xu}}, \bibinfo {author}
  {\bibfnamefont {S.}~\bibnamefont {Bubin}}, \bibinfo {author} {\bibfnamefont
  {E.}~\bibnamefont {L\"otstedt}}, \bibinfo {author} {\bibfnamefont
  {S.}~\bibnamefont {Erattuphuza}}, \bibinfo {author} {\bibfnamefont
  {A.}~\bibnamefont {Iwasaki}}, \bibinfo {author} {\bibfnamefont
  {D.}~\bibnamefont {Kartashov}}, \bibinfo {author} {\bibfnamefont
  {K.}~\bibnamefont {Varga}}, \bibinfo {author} {\bibfnamefont
  {G.}~\bibnamefont {G.~Paulus}}, \bibinfo {author} {\bibfnamefont
  {A.}~\bibnamefont {Baltu\ifmmode~\check{s}\else \v{s}\fi{}ka}}, \bibinfo
  {author} {\bibfnamefont {K.}~\bibnamefont {Yamanouchi}}, \ and\ \bibinfo
  {author} {\bibfnamefont {M.}~\bibnamefont {Kitzler}},\ }\href {\doibase
  10.1103/PhysRevA.89.023429} {\bibfield  {journal} {\bibinfo  {journal} {Phys.
  Rev. A}\ }\textbf {\bibinfo {volume} {89}},\ \bibinfo {pages} {023429}
  (\bibinfo {year} {2014})}\BibitemShut {NoStop}%
\bibitem [{\citenamefont {Cornaggia}(2016)}]{Cornaggia.JPB.2016}%
  \BibitemOpen
  \bibfield  {author} {\bibinfo {author} {\bibfnamefont {C.}~\bibnamefont
  {Cornaggia}},\ }\href {http://stacks.iop.org/0953-4075/49/i=19/a=19LT01}
  {\bibfield  {journal} {\bibinfo  {journal} {J. Phys. B: At. Mol. Opt. Phys.}\
  }\textbf {\bibinfo {volume} {49}},\ \bibinfo {pages} {19LT01} (\bibinfo
  {year} {2016})}\BibitemShut {NoStop}%
\bibitem [{\citenamefont {Liu}\ and\ \citenamefont
  {Barth}(2017)}]{Liu.PRL.2017}%
  \BibitemOpen
  \bibfield  {author} {\bibinfo {author} {\bibfnamefont {K.}~\bibnamefont
  {Liu}}\ and\ \bibinfo {author} {\bibfnamefont {I.}~\bibnamefont {Barth}},\
  }\href {\doibase 10.1103/PhysRevLett.119.243204} {\bibfield  {journal}
  {\bibinfo  {journal} {Phys. Rev. Lett.}\ }\textbf {\bibinfo {volume} {119}},\
  \bibinfo {pages} {243204} (\bibinfo {year} {2017})}\BibitemShut {NoStop}%
\bibitem [{\citenamefont {Plummer}\ and\ \citenamefont
  {McCann}(1996)}]{Plummer1996}%
  \BibitemOpen
  \bibfield  {author} {\bibinfo {author} {\bibfnamefont {M.}~\bibnamefont
  {Plummer}}\ and\ \bibinfo {author} {\bibfnamefont {J.~F.}\ \bibnamefont
  {McCann}},\ }\href@noop {} {\bibfield  {journal} {\bibinfo  {journal} {J.
  Phys. B: At. Mol. Opt. Phys.}\ }\textbf {\bibinfo {volume} {29}},\ \bibinfo
  {pages} {4625} (\bibinfo {year} {1996})}\BibitemShut {NoStop}%
\bibitem [{\citenamefont {Plummer}\ and\ \citenamefont
  {McCann}(1997)}]{Plummer1997}%
  \BibitemOpen
  \bibfield  {author} {\bibinfo {author} {\bibfnamefont {M.}~\bibnamefont
  {Plummer}}\ and\ \bibinfo {author} {\bibfnamefont {J.~F.}\ \bibnamefont
  {McCann}},\ }\href@noop {} {\bibfield  {journal} {\bibinfo  {journal} {J.
  Phys. B: At. Mol. Opt. Phys.}\ }\textbf {\bibinfo {volume} {30}},\ \bibinfo
  {pages} {L401} (\bibinfo {year} {1997})}\BibitemShut {NoStop}%
\bibitem [{\citenamefont {Takemoto}\ and\ \citenamefont
  {Becker}(2010)}]{Takemoto2010}%
  \BibitemOpen
  \bibfield  {author} {\bibinfo {author} {\bibfnamefont {N.}~\bibnamefont
  {Takemoto}}\ and\ \bibinfo {author} {\bibfnamefont {A.}~\bibnamefont
  {Becker}},\ }\href {\doibase 10.1103/PhysRevLett.105.203004} {\bibfield
  {journal} {\bibinfo  {journal} {Phys. Rev. Lett.}\ }\textbf {\bibinfo
  {volume} {105}},\ \bibinfo {pages} {203004} (\bibinfo {year}
  {2010})}\BibitemShut {NoStop}%
\bibitem [{\citenamefont {Odenweller}\ \emph {et~al.}(2011)\citenamefont
  {Odenweller}, \citenamefont {Takemoto}, \citenamefont {Vredenborg},
  \citenamefont {Cole}, \citenamefont {Pahl}, \citenamefont {Titze},
  \citenamefont {Schmidt}, \citenamefont {Jahnke}, \citenamefont {D\"orner},\
  and\ \citenamefont {Becker}}]{Odenweller2011}%
  \BibitemOpen
  \bibfield  {author} {\bibinfo {author} {\bibfnamefont {M.}~\bibnamefont
  {Odenweller}}, \bibinfo {author} {\bibfnamefont {N.}~\bibnamefont
  {Takemoto}}, \bibinfo {author} {\bibfnamefont {A.}~\bibnamefont
  {Vredenborg}}, \bibinfo {author} {\bibfnamefont {K.}~\bibnamefont {Cole}},
  \bibinfo {author} {\bibfnamefont {K.}~\bibnamefont {Pahl}}, \bibinfo {author}
  {\bibfnamefont {J.}~\bibnamefont {Titze}}, \bibinfo {author} {\bibfnamefont
  {L.~P.~H.}\ \bibnamefont {Schmidt}}, \bibinfo {author} {\bibfnamefont
  {T.}~\bibnamefont {Jahnke}}, \bibinfo {author} {\bibfnamefont
  {R.}~\bibnamefont {D\"orner}}, \ and\ \bibinfo {author} {\bibfnamefont
  {A.}~\bibnamefont {Becker}},\ }\href {\doibase
  10.1103/PhysRevLett.107.143004} {\bibfield  {journal} {\bibinfo  {journal}
  {Phys. Rev. Lett.}\ }\textbf {\bibinfo {volume} {107}},\ \bibinfo {pages}
  {143004} (\bibinfo {year} {2011})}\BibitemShut {NoStop}%
\bibitem [{\citenamefont {Silva}\ \emph {et~al.}(2013)\citenamefont {Silva},
  \citenamefont {Catoire}, \citenamefont {Rivi\`ere}, \citenamefont {Bachau},\
  and\ \citenamefont {Mart\'{\i}n}}]{Silva2013a}%
  \BibitemOpen
  \bibfield  {author} {\bibinfo {author} {\bibfnamefont {R.~E.~F.}\
  \bibnamefont {Silva}}, \bibinfo {author} {\bibfnamefont {F.}~\bibnamefont
  {Catoire}}, \bibinfo {author} {\bibfnamefont {P.}~\bibnamefont {Rivi\`ere}},
  \bibinfo {author} {\bibfnamefont {H.}~\bibnamefont {Bachau}}, \ and\ \bibinfo
  {author} {\bibfnamefont {F.}~\bibnamefont {Mart\'{\i}n}},\ }\href {\doibase
  10.1103/PhysRevLett.110.113001} {\bibfield  {journal} {\bibinfo  {journal}
  {Phys. Rev. Lett.}\ }\textbf {\bibinfo {volume} {110}},\ \bibinfo {pages}
  {113001} (\bibinfo {year} {2013})}\BibitemShut {NoStop}%
\bibitem [{\citenamefont {Yue}\ and\ \citenamefont {Madsen}(2016)}]{Yue2016b}%
  \BibitemOpen
  \bibfield  {author} {\bibinfo {author} {\bibfnamefont {L.}~\bibnamefont
  {Yue}}\ and\ \bibinfo {author} {\bibfnamefont {L.~B.}\ \bibnamefont
  {Madsen}},\ }\href {\doibase 10.1103/PhysRevA.93.031401} {\bibfield
  {journal} {\bibinfo  {journal} {Phys. Rev. A}\ }\textbf {\bibinfo {volume}
  {93}},\ \bibinfo {pages} {031401} (\bibinfo {year} {2016})}\BibitemShut
  {NoStop}%
\bibitem [{\citenamefont {Xu}\ \emph {et~al.}(2017)\citenamefont {Xu},
  \citenamefont {Li}, \citenamefont {He}, \citenamefont {Wang}, \citenamefont
  {Kielpinski}, \citenamefont {Sang},\ and\ \citenamefont
  {Litvinyuk}}]{Xu2017}%
  \BibitemOpen
  \bibfield  {author} {\bibinfo {author} {\bibfnamefont {H.}~\bibnamefont
  {Xu}}, \bibinfo {author} {\bibfnamefont {Z.}~\bibnamefont {Li}}, \bibinfo
  {author} {\bibfnamefont {F.}~\bibnamefont {He}}, \bibinfo {author}
  {\bibfnamefont {X.}~\bibnamefont {Wang}}, \bibinfo {author} {\bibfnamefont
  {D.}~\bibnamefont {Kielpinski}}, \bibinfo {author} {\bibfnamefont {R.~T.}\
  \bibnamefont {Sang}}, \ and\ \bibinfo {author} {\bibfnamefont {I.~V.}\
  \bibnamefont {Litvinyuk}},\ }\href {\doibase 10.1038/ncomms15849} {\bibfield
  {journal} {\bibinfo  {journal} {Nat. Comm.}\ }\textbf {\bibinfo {volume}
  {8}},\ \bibinfo {pages} {1} (\bibinfo {year} {2017})}\BibitemShut {NoStop}%
\bibitem [{\citenamefont {Ben-Itzhak}\ \emph {et~al.}(2005)\citenamefont
  {Ben-Itzhak}, \citenamefont {Wang}, \citenamefont {Xia}, \citenamefont
  {Sayler}, \citenamefont {Smith}, \citenamefont {Carnes},\ and\ \citenamefont
  {Esry}}]{Ben-Itzhak2005}%
  \BibitemOpen
  \bibfield  {author} {\bibinfo {author} {\bibfnamefont {I.}~\bibnamefont
  {Ben-Itzhak}}, \bibinfo {author} {\bibfnamefont {P.~Q.}\ \bibnamefont
  {Wang}}, \bibinfo {author} {\bibfnamefont {J.~F.}\ \bibnamefont {Xia}},
  \bibinfo {author} {\bibfnamefont {A.~M.}\ \bibnamefont {Sayler}}, \bibinfo
  {author} {\bibfnamefont {M.~A.}\ \bibnamefont {Smith}}, \bibinfo {author}
  {\bibfnamefont {K.~D.}\ \bibnamefont {Carnes}}, \ and\ \bibinfo {author}
  {\bibfnamefont {B.~D.}\ \bibnamefont {Esry}},\ }\href {\doibase
  10.1103/PhysRevLett.95.073002} {\bibfield  {journal} {\bibinfo  {journal}
  {Phys. Rev. Lett.}\ }\textbf {\bibinfo {volume} {95}},\ \bibinfo {pages}
  {073002} (\bibinfo {year} {2005})}\BibitemShut {NoStop}%
\bibitem [{\citenamefont {Rathje}\ \emph {et~al.}(2013)\citenamefont {Rathje},
  \citenamefont {Sayler}, \citenamefont {Zeng}, \citenamefont {Wustelt},
  \citenamefont {Figger}, \citenamefont {Esry},\ and\ \citenamefont
  {Paulus}}]{Rathje2013}%
  \BibitemOpen
  \bibfield  {author} {\bibinfo {author} {\bibfnamefont {T.}~\bibnamefont
  {Rathje}}, \bibinfo {author} {\bibfnamefont {A.~M.}\ \bibnamefont {Sayler}},
  \bibinfo {author} {\bibfnamefont {S.}~\bibnamefont {Zeng}}, \bibinfo {author}
  {\bibfnamefont {P.}~\bibnamefont {Wustelt}}, \bibinfo {author} {\bibfnamefont
  {H.}~\bibnamefont {Figger}}, \bibinfo {author} {\bibfnamefont {B.~D.}\
  \bibnamefont {Esry}}, \ and\ \bibinfo {author} {\bibfnamefont {G.~G.}\
  \bibnamefont {Paulus}},\ }\href {\doibase 10.1103/PhysRevLett.111.093002}
  {\bibfield  {journal} {\bibinfo  {journal} {Phys. Rev. Lett.}\ }\textbf
  {\bibinfo {volume} {111}},\ \bibinfo {pages} {093002} (\bibinfo {year}
  {2013})}\BibitemShut {NoStop}%
\bibitem [{\citenamefont {Wustelt}\ \emph {et~al.}(2015)\citenamefont
  {Wustelt}, \citenamefont {M\"oller}, \citenamefont {Rathje}, \citenamefont
  {Sayler}, \citenamefont {St\"ohlker},\ and\ \citenamefont
  {Paulus}}]{Wustelt2015}%
  \BibitemOpen
  \bibfield  {author} {\bibinfo {author} {\bibfnamefont {P.}~\bibnamefont
  {Wustelt}}, \bibinfo {author} {\bibfnamefont {M.}~\bibnamefont {M\"oller}},
  \bibinfo {author} {\bibfnamefont {T.}~\bibnamefont {Rathje}}, \bibinfo
  {author} {\bibfnamefont {A.~M.}\ \bibnamefont {Sayler}}, \bibinfo {author}
  {\bibfnamefont {T.}~\bibnamefont {St\"ohlker}}, \ and\ \bibinfo {author}
  {\bibfnamefont {G.~G.}\ \bibnamefont {Paulus}},\ }\href {\doibase
  10.1103/PhysRevA.91.031401} {\bibfield  {journal} {\bibinfo  {journal} {Phys.
  Rev. A}\ }\textbf {\bibinfo {volume} {91}},\ \bibinfo {pages} {031401}
  (\bibinfo {year} {2015})}\BibitemShut {NoStop}%
\bibitem [{Note1()}]{Note1}%
  \BibitemOpen
  \bibinfo {note} {See Supplemental Material at [URL will be inserted by
  publisher] for a PDF document further detailing the experimental methods and
  numerical simulations.}\BibitemShut {Stop}%
\bibitem [{\citenamefont {Odenweller}\ \emph {et~al.}(2014)\citenamefont
  {Odenweller}, \citenamefont {Lower}, \citenamefont {Pahl}, \citenamefont
  {Sch\"utt}, \citenamefont {Wu}, \citenamefont {Cole}, \citenamefont
  {Vredenborg}, \citenamefont {Schmidt}, \citenamefont {Neumann}, \citenamefont
  {Titze}, \citenamefont {Jahnke}, \citenamefont {Meckel}, \citenamefont
  {Kunitski}, \citenamefont {Havermeier}, \citenamefont {Voss}, \citenamefont
  {Sch\"offler}, \citenamefont {Sann}, \citenamefont {Voigtsberger},
  \citenamefont {Schmidt-B\"ocking},\ and\ \citenamefont
  {D\"orner}}]{Odenweller2014}%
  \BibitemOpen
  \bibfield  {author} {\bibinfo {author} {\bibfnamefont {M.}~\bibnamefont
  {Odenweller}}, \bibinfo {author} {\bibfnamefont {J.}~\bibnamefont {Lower}},
  \bibinfo {author} {\bibfnamefont {K.}~\bibnamefont {Pahl}}, \bibinfo {author}
  {\bibfnamefont {M.}~\bibnamefont {Sch\"utt}}, \bibinfo {author}
  {\bibfnamefont {J.}~\bibnamefont {Wu}}, \bibinfo {author} {\bibfnamefont
  {K.}~\bibnamefont {Cole}}, \bibinfo {author} {\bibfnamefont {A.}~\bibnamefont
  {Vredenborg}}, \bibinfo {author} {\bibfnamefont {L.~P.}\ \bibnamefont
  {Schmidt}}, \bibinfo {author} {\bibfnamefont {N.}~\bibnamefont {Neumann}},
  \bibinfo {author} {\bibfnamefont {J.}~\bibnamefont {Titze}}, \bibinfo
  {author} {\bibfnamefont {T.}~\bibnamefont {Jahnke}}, \bibinfo {author}
  {\bibfnamefont {M.}~\bibnamefont {Meckel}}, \bibinfo {author} {\bibfnamefont
  {M.}~\bibnamefont {Kunitski}}, \bibinfo {author} {\bibfnamefont
  {T.}~\bibnamefont {Havermeier}}, \bibinfo {author} {\bibfnamefont
  {S.}~\bibnamefont {Voss}}, \bibinfo {author} {\bibfnamefont {M.}~\bibnamefont
  {Sch\"offler}}, \bibinfo {author} {\bibfnamefont {H.}~\bibnamefont {Sann}},
  \bibinfo {author} {\bibfnamefont {J.}~\bibnamefont {Voigtsberger}}, \bibinfo
  {author} {\bibfnamefont {H.}~\bibnamefont {Schmidt-B\"ocking}}, \ and\
  \bibinfo {author} {\bibfnamefont {R.}~\bibnamefont {D\"orner}},\ }\href
  {\doibase 10.1103/PhysRevA.89.013424} {\bibfield  {journal} {\bibinfo
  {journal} {Phys. Rev. A}\ }\textbf {\bibinfo {volume} {89}},\ \bibinfo
  {pages} {013424} (\bibinfo {year} {2014})}\BibitemShut {NoStop}%
\bibitem [{\citenamefont {Yue}\ and\ \citenamefont {Madsen}(2013)}]{Yue2013}%
  \BibitemOpen
  \bibfield  {author} {\bibinfo {author} {\bibfnamefont {L.}~\bibnamefont
  {Yue}}\ and\ \bibinfo {author} {\bibfnamefont {L.~B.}\ \bibnamefont
  {Madsen}},\ }\href {\doibase 10.1103/PhysRevA.88.063420} {\bibfield
  {journal} {\bibinfo  {journal} {Phys. Rev. A}\ }\textbf {\bibinfo {volume}
  {88}},\ \bibinfo {pages} {063420} (\bibinfo {year} {2013})}\BibitemShut
  {NoStop}%
\bibitem [{\citenamefont {Wu}\ \emph {et~al.}(2013)\citenamefont {Wu},
  \citenamefont {Kunitski}, \citenamefont {Pitzer}, \citenamefont {Trinter},
  \citenamefont {Schmidt}, \citenamefont {Jahnke}, \citenamefont
  {Magrakvelidze}, \citenamefont {Madsen}, \citenamefont {Madsen},
  \citenamefont {Thumm},\ and\ \citenamefont {D\"orner}}]{Wu2013}%
  \BibitemOpen
  \bibfield  {author} {\bibinfo {author} {\bibfnamefont {J.}~\bibnamefont
  {Wu}}, \bibinfo {author} {\bibfnamefont {M.}~\bibnamefont {Kunitski}},
  \bibinfo {author} {\bibfnamefont {M.}~\bibnamefont {Pitzer}}, \bibinfo
  {author} {\bibfnamefont {F.}~\bibnamefont {Trinter}}, \bibinfo {author}
  {\bibfnamefont {L.~P.~H.}\ \bibnamefont {Schmidt}}, \bibinfo {author}
  {\bibfnamefont {T.}~\bibnamefont {Jahnke}}, \bibinfo {author} {\bibfnamefont
  {M.}~\bibnamefont {Magrakvelidze}}, \bibinfo {author} {\bibfnamefont {C.~B.}\
  \bibnamefont {Madsen}}, \bibinfo {author} {\bibfnamefont {L.~B.}\
  \bibnamefont {Madsen}}, \bibinfo {author} {\bibfnamefont {U.}~\bibnamefont
  {Thumm}}, \ and\ \bibinfo {author} {\bibfnamefont {R.}~\bibnamefont
  {D\"orner}},\ }\href {\doibase 10.1103/PhysRevLett.111.023002} {\bibfield
  {journal} {\bibinfo  {journal} {Phys. Rev. Lett.}\ }\textbf {\bibinfo
  {volume} {111}},\ \bibinfo {pages} {023002} (\bibinfo {year}
  {2013})}\BibitemShut {NoStop}%
\bibitem [{\citenamefont {Takemoto}\ and\ \citenamefont
  {Becker}(2011)}]{Takemoto2011}%
  \BibitemOpen
  \bibfield  {author} {\bibinfo {author} {\bibfnamefont {N.}~\bibnamefont
  {Takemoto}}\ and\ \bibinfo {author} {\bibfnamefont {A.}~\bibnamefont
  {Becker}},\ }\href {\doibase 10.1103/PhysRevA.84.023401} {\bibfield
  {journal} {\bibinfo  {journal} {Phys. Rev. A}\ }\textbf {\bibinfo {volume}
  {84}},\ \bibinfo {pages} {023401} (\bibinfo {year} {2011})}\BibitemShut
  {NoStop}%
\bibitem [{\citenamefont {Becker}\ \emph {et~al.}(2002)\citenamefont {Becker},
  \citenamefont {Grasbon}, \citenamefont {Kopold}, \citenamefont
  {Milo{\v{s}}evi{\'{c}}},\ and\ \citenamefont {Walther}}]{Becker2002}%
  \BibitemOpen
  \bibfield  {author} {\bibinfo {author} {\bibfnamefont {W.}~\bibnamefont
  {Becker}}, \bibinfo {author} {\bibfnamefont {F.}~\bibnamefont {Grasbon}},
  \bibinfo {author} {\bibfnamefont {R.}~\bibnamefont {Kopold}}, \bibinfo
  {author} {\bibfnamefont {D.~B.}\ \bibnamefont {Milo{\v{s}}evi{\'{c}}}}, \
  and\ \bibinfo {author} {\bibfnamefont {H.}~\bibnamefont {Walther}},\ }in\
  \href {\doibase https://doi.org/10.1016/S1049-250X(02)80006-4} {\emph
  {\bibinfo {booktitle} {Advances in Atomic Molecular and Optical Physics}}},\
  \bibinfo {series} {Advances In Atomic, Molecular, and Optical Physics},
  Vol.~\bibinfo {volume} {48},\ \bibinfo {editor} {edited by\ \bibinfo {editor}
  {\bibfnamefont {B.}~\bibnamefont {Bederson}}\ and\ \bibinfo {editor}
  {\bibfnamefont {H.}~\bibnamefont {Walther}}}\ (\bibinfo  {publisher}
  {Academic Press},\ \bibinfo {address} {525 B St., suite 1900, San Diego, CA
  92101-4495 USA},\ \bibinfo {year} {2002})\ pp.\ \bibinfo {pages} {35 --
  98}\BibitemShut {NoStop}%
\bibitem [{\citenamefont {Schwendner}\ \emph {et~al.}(1997)\citenamefont
  {Schwendner}, \citenamefont {Seyl},\ and\ \citenamefont
  {Schinke}}]{Schwendner1997}%
  \BibitemOpen
  \bibfield  {author} {\bibinfo {author} {\bibfnamefont {P.}~\bibnamefont
  {Schwendner}}, \bibinfo {author} {\bibfnamefont {F.}~\bibnamefont {Seyl}}, \
  and\ \bibinfo {author} {\bibfnamefont {R.}~\bibnamefont {Schinke}},\ }\href
  {\doibase 10.1016/S0301-0104(97)00045-1} {\bibfield  {journal} {\bibinfo
  {journal} {Chem. Phys.}\ }\textbf {\bibinfo {volume} {217}},\ \bibinfo
  {pages} {233} (\bibinfo {year} {1997})}\BibitemShut {NoStop}%
\bibitem [{\citenamefont {Staudte}\ \emph {et~al.}(2007)\citenamefont
  {Staudte}, \citenamefont {Pavi\ifmmode \check{c}\else
  \v{c}\fi{}i\ifmmode~\acute{c}\else \'{c}\fi{}}, \citenamefont {Chelkowski},
  \citenamefont {Zeidler}, \citenamefont {Meckel}, \citenamefont {Niikura},
  \citenamefont {Sch\"offler}, \citenamefont {Sch\"ossler}, \citenamefont
  {Ulrich}, \citenamefont {Rajeev}, \citenamefont {Weber}, \citenamefont
  {Jahnke}, \citenamefont {Villeneuve}, \citenamefont {Bandrauk}, \citenamefont
  {Cocke}, \citenamefont {Corkum},\ and\ \citenamefont
  {D\"orner}}]{Staudte2007}%
  \BibitemOpen
  \bibfield  {author} {\bibinfo {author} {\bibfnamefont {A.}~\bibnamefont
  {Staudte}}, \bibinfo {author} {\bibfnamefont {D.}~\bibnamefont {Pavi\ifmmode
  \check{c}\else \v{c}\fi{}i\ifmmode~\acute{c}\else \'{c}\fi{}}}, \bibinfo
  {author} {\bibfnamefont {S.}~\bibnamefont {Chelkowski}}, \bibinfo {author}
  {\bibfnamefont {D.}~\bibnamefont {Zeidler}}, \bibinfo {author} {\bibfnamefont
  {M.}~\bibnamefont {Meckel}}, \bibinfo {author} {\bibfnamefont
  {H.}~\bibnamefont {Niikura}}, \bibinfo {author} {\bibfnamefont
  {M.}~\bibnamefont {Sch\"offler}}, \bibinfo {author} {\bibfnamefont
  {S.}~\bibnamefont {Sch\"ossler}}, \bibinfo {author} {\bibfnamefont
  {B.}~\bibnamefont {Ulrich}}, \bibinfo {author} {\bibfnamefont {P.~P.}\
  \bibnamefont {Rajeev}}, \bibinfo {author} {\bibfnamefont {T.}~\bibnamefont
  {Weber}}, \bibinfo {author} {\bibfnamefont {T.}~\bibnamefont {Jahnke}},
  \bibinfo {author} {\bibfnamefont {D.~M.}\ \bibnamefont {Villeneuve}},
  \bibinfo {author} {\bibfnamefont {A.~D.}\ \bibnamefont {Bandrauk}}, \bibinfo
  {author} {\bibfnamefont {C.~L.}\ \bibnamefont {Cocke}}, \bibinfo {author}
  {\bibfnamefont {P.~B.}\ \bibnamefont {Corkum}}, \ and\ \bibinfo {author}
  {\bibfnamefont {R.}~\bibnamefont {D\"orner}},\ }\href {\doibase
  10.1103/PhysRevLett.98.073003} {\bibfield  {journal} {\bibinfo  {journal}
  {Phys. Rev. Lett.}\ }\textbf {\bibinfo {volume} {98}},\ \bibinfo {pages}
  {073003} (\bibinfo {year} {2007})}\BibitemShut {NoStop}%
\end{thebibliography}
\end{document}


\title{Supplemental Material: Goldilocks Zone for Enhanced Ionization in Strong Laser Fields}

\author{M. M\"oller}
\affiliation{Institute of Optics and Quantum Electronics, Abbe Center of Photonics, Friedrich Schiller University Jena, Max-Wien-Platz 1, 07743 Jena, Germany}
\affiliation{Helmholtz Institut Jena, Fr\"obelstieg 3, 07743 Jena, Germany}

\author{A. M. Sayler}
\affiliation{Institute of Optics and Quantum Electronics, Abbe Center of Photonics, Friedrich Schiller University Jena, Max-Wien-Platz 1, 07743 Jena, Germany}
\affiliation{Helmholtz Institut Jena, Fr\"obelstieg 3, 07743 Jena, Germany}

\author{P. Wustelt}
\affiliation{Institute of Optics and Quantum Electronics, Abbe Center of Photonics, Friedrich Schiller University Jena, Max-Wien-Platz 1, 07743 Jena, Germany}
\affiliation{Helmholtz Institut Jena, Fr\"obelstieg 3, 07743 Jena, Germany}

\author{L. Yue}
\affiliation{Institute of Physical Chemistry and Abbe Center of Photonics, Friedrich Schiller University Jena, Helmholtzweg 4, 07743 Jena, Germany}

\author{S. Gr\"afe}
\affiliation{Institute of Physical Chemistry and Abbe Center of Photonics, Friedrich Schiller University Jena, Helmholtzweg 4, 07743 Jena, Germany}

\author{G. G. Paulus}
\affiliation{Institute of Optics and Quantum Electronics, Abbe Center of Photonics, Friedrich Schiller University Jena, Max-Wien-Platz 1, 07743 Jena, Germany}
\affiliation{Helmholtz Institut Jena, Fr\"obelstieg 3, 07743 Jena, Germany}

\maketitle

\section*{Experimental details}

The measurements  here were done with a highly automated ion-beam target coincidence 3D momentum imaging setup \cite{Ben-Itzhak2005, Rathje2013, Wustelt2015}. The H$_2^+$ ion beam is created in a duoplasmatron ion source, accelerated to 9 keV, velocity and mass selected with a Wien filter and collimated using a series of Einzel lenses, adjustable apertures and electrostatic deflectors. Additionally, a longitudinal electric field in the interaction region is used to distinguish charged and neutral fragments and allow coincidence detection of all fragmentation channels using a  time- and position-sensitive multi-channel plate (MCP) delay-line detector (RoentDek DLD). 

Here we choose to work in the SWIR ($\lambda = 2~\mu$m), not the much more common Ti:Sapphire IR regime ($\lambda\sim 800$ nm), to enable determination of the electron momentum, $\vec{p}_{e}$, from the shift in the center-of-mass of the detected protons, i.e.~$\vec{p}_{e}=-(\vec{p}_{H^+}+\vec{p}_{H^+})$. In addition to a Ti:Sapphire tabletop system (Femtopower Compact Pro, Femtolasers) and a high power booster amplification stage (Thales), this requires operation of a high-energy, high-average-power optic parametric amplifier (HE-TOPAS OPA, Light Conversion) to achieve the 65 fs (full-width at half-maximum in intensity), 2 $\mu$m, 750 $\mu$J pulses. The pulses are linearly polarized using an acute angle reflection off a Germanium plate and focused with an $f=150$ mm, 90$^\circ$ off-axis parabolic mirror to peak intensities up to 1.3 PW/cm$^2$ at 1 kHz. In addition, we continuously scan the intensity and tag each fragmentation event with that information, which allows for subsequent sorting and a detailed investigation of the intensity dependence. 

\section*{Wavepacket propagation}

The bound electron-nuclear wavefunction of H$_2^+$ is approximated using solutions to the time-dependent Schr\"odinger equation for the two lowest field-free Born-Oppenheimer (BO) electronic states \cite{Schwendner1997, Staudte2007}. The external laser field, $E(t)$, couples the electronic states with the transition dipole moment, $d_{gu}(R)$, and facilitates the laser-driven nuclear dynamics, i.e.~stretching of the internuclear distance due to transitions of the nuclear wavepacket between the $1s\sigma_g$ and $2p\sigma_u$ states.
\begin{equation}
i\frac{d}{dt}
\begin{bmatrix}
    \Psi_{1s\sigma_g}(R,t)  \\
    \Psi_{2p\sigma_u}(R,t) \\
\end{bmatrix}=
\begin{bmatrix}
   V_{1s\sigma_g}(R)+\frac{p^2_R}{2\mu}~~~ & -d_{gu}(R)\cdot E(t)  \\
    -d_{gu}(R)\cdot E(t) ~~~ & V_{2p\sigma_u}(R)+\frac{p^2_R}{2\mu} \\
\end{bmatrix}
\begin{bmatrix}
    \Psi_{1s\sigma_g}(R,t)  \\
    \Psi_{2p\sigma_u}(R,t) \\
\end{bmatrix}
\label{eq.tdse}
\end{equation}
Ionization, i.e.~the transition to the $1/R$ potential curve from the perspective of the nuclei, is incorporated at each time-step by the use of $R$- and field-strength-dependent ionization rates for both of the electronic states, i.e.~$\Gamma_g(R,|E(t)|)$ and $\Gamma_u(R,|E(t)|)$ \cite{Plummer1996,Plummer1997}.
Namely, since the probability decays as $|\Psi'|^2=|\Psi|^2\exp(-\Gamma dt)$
\begin{equation}
\begin{bmatrix}
    \Psi_{1s\sigma_g}(R,t+dt)  \\
    \Psi_{2p\sigma_u}(R,t+dt) \\
\end{bmatrix}=
\begin{bmatrix}
   \exp(-\frac{\Gamma_g(R,|E(t)|)\cdot dt}{2}) & 0  \\
    0 &    \exp(-\frac{\Gamma_u(R,|E(t)|)\cdot dt}{2}) \\
\end{bmatrix}
\begin{bmatrix}
    \Psi_{1s\sigma_g}(R,t)  \\
    \Psi_{2p\sigma_u}(R,t) \\
\end{bmatrix}
\label{eq.tdse2}
\end{equation}

The calculation is done using the split-step method to solve the TDSE, Eq.~\ref{eq.tdse}, within a box defined from $R=0$ to 50 a.u.~with a spatial grid of $dR=0.025$ a.u.~and a time step of $dt=0.1$ a.u.~covering the $\cos^2$ envelope of the electric field. We have checked that the nuclear momentum and ionization probabilities converge for these parameters. 

This method allows one to track the nuclear wavefunction and determine the $R$- and $t$-dependent ionization yield, $P_{\textrm{ion}}(R,t)$, i.e.
\begin{equation}
P_{\textrm{ion}}(R,t) = (|\Psi_{1s\sigma_g}(R,t)|^2-|\Psi_{1s\sigma_g}(R,t+dt)|^2)+
(|\Psi_{2p\sigma_u}(R,t)|^2-|\Psi_{2p\sigma_u}(R,t+dt)|^2)~~,
\label{eq.tdse4}
\end{equation}
using the values from Eq.~\ref{eq.tdse2}.
 This can then be converted into a nuclear-energy- and electron-energy-dependent probability, i.e.~$P(\textrm{KER},E_e)$, using the approximate nuclear energy, $E_{\textrm{KER}_i} \approx 1/R_{i}$, and the approximate electron energy, $p_{e_i}\approx-A(t_{i})$, based on electric field's vector potential, $A(t)=-\int_{-\infty}^t E(t')dt'$.
Additionally, to match the experimental conditions, the theoretical results are averaged over the intensity distribution in the focal volume and the incoherent sum of vibrational states, $\nu=0$ -- 15, described by the Franck-Condon approximation, is used.

\section*{Intensity envelope optimization}
To produce the characteristic double-peak structure of EI, the laser intensity must ramp up in a very particular way, see Fig.~4(a)--(c). Namely, the molecular dissociation is initiated by the laser field at relatively low intensities, $I_{\textrm{diss}}$, then ionization will occur at a later time, $t_{\textrm{ion}} = t_{\textrm{diss}} + \Delta t$, after the molecule has stretched to $R_{\textrm{ion}}$ and the intensity has increased sufficiently to ionize the molecule, $I_{\textrm{ion}}$.
Thus, for a  cos$^2$ pulse,  $I(t,\textrm{FWHM}) = I_0 \cos\bigl(t\frac{\pi}{2\cdot\textrm{FWHM}}\bigr)^2$, the optimal intensity-dependent pulse duration for a given peak intensity, $I_0$, is
\begin{equation}
\textrm{FWHM}(\Delta t,I_{\textrm{ion}},I_{\textrm{dis}};I_0) = \frac{\Delta t}{\frac{2}{\pi}\bigl[\cos^{-1}(\sqrt{I_{\textrm{diss}}/I_0})-\cos^{-1}(\sqrt{I_{\textrm{ion}}/I_0})\bigr]}
\label{fit}
\end{equation}

To determine the optimal parameters, one can use the measured and calculated position of the peak contrast values, see Fig.~4(e) (circles), to fit
Eq.~\ref{fit} and determine the parameters: $\Delta t$, $I_{\textrm{ion}}$, and $I_{\textrm{dis}}$.  For the intensity averaged data, this yields the fit parameters: $I_{\textrm{diss}}\simeq 5.8\cdot 10^{13}~\textrm{W/cm}^2$, $I_{\textrm{ion}}\simeq 4.2\cdot 10^{14}~\textrm{W/cm}^2$ and $\Delta t \simeq 20~\textrm{fs}$, see Fig.~\ref{fig:4}(e) (line). For the single intensity calculations, this yields the fit parameters: $I_{\textrm{diss}}\simeq 1.4\cdot 10^{14}~\textrm{W/cm}^2$, $I_{\textrm{ion}}\simeq 2.9\cdot 10^{14}~\textrm{W/cm}^2$ and $\Delta t \simeq 20~\textrm{fs}$. \ifWC (80) \fi


%